\documentclass[conference]{IEEEtran}
\IEEEoverridecommandlockouts

\usepackage[T1]{fontenc}
\usepackage[utf8]{inputenc}
\usepackage{cite}
\usepackage{amsmath,amssymb,amsfonts}
\usepackage{graphicx}
\usepackage{textcomp}
\usepackage{xcolor}
\usepackage{booktabs}
\usepackage{array}
\usepackage{url}
\usepackage{hyperref}
\hypersetup{hidelinks,colorlinks=false}

\def\BibTeX{{\rm B\kern-.05em{\sc i\kern-.025em b}\kern-.08em
    T\kern-.1667em\lower.7ex\hbox{E}\kern-.125emX}}

\begin{document}

\title{Deploy, Calibrate, Monitor, Heal --- No Human Required:\\
An Autonomous AI SRE Agent for Elasticsearch}

\author{\IEEEauthorblockN{Muhamed Ramees Cheriya Mukkolakkal}
\IEEEauthorblockA{\textit{rameescm1@gmail.com}\\
March 2026 $\cdot$ Cluster: Elasticsearch 8.17.0, 3 master + 12 data nodes, Kubernetes}
}

\maketitle

\begin{abstract}
Operating Elasticsearch clusters at scale demands continuous human expertise spanning the full lifecycle --- from initial deployment through performance tuning, monitoring, failure prediction, and incident recovery. We present the ES Guardian Agent, an autonomous AI SRE system that manages the complete Elasticsearch lifecycle without human intervention through eleven distinct phases: Evaluate, Optimize, Deploy, Calibrate, Stabilize, Alert, Predict, Heal, Learn, and Upgrade. A critical differentiator of the Guardian Agent is its multi-source predictive failure engine. The system continuously ingests and correlates metrics trends, application logs, and kernel-level telemetry --- including Linux dmesg streams, NVMe SMART data, NIC bond statistics, and thermal sensors --- to anticipate failures hours before they materialize. By cross-referencing current system signatures against a persistent incident memory of resolved failures, the AI engine prepares detailed remediation playbooks in advance and stages corrective actions proactively. This predictive posture, rather than reactive response, is the architectural cornerstone enabling six-nines (99.9999\%) availability targets --- where even minutes of unplanned downtime per year are unacceptable. Through four successive agent architectures --- culminating in a 4,589-line system with five monitoring layers and an iterative AI action loop --- we demonstrate that an LLM equipped with tool-use access can function as a full-lifecycle autonomous SRE. In production evaluation, the Guardian Agent executed 300 autonomous investigation-and-repair cycles, recovered a cluster from an 18-hour cross-system outage, diagnosed hardware NIC failures across all host nodes, and maintained continuous operational visibility. We establish that data volume per shard --- not tuning --- is the primary determinant of query performance, with latency scaling at 0.26~ms per MB/shard.
\end{abstract}


\section{Introduction}

\subsection{Motivation}

Modern Elasticsearch deployments face operational challenges spanning the full system stack --- from hardware failures and host disk pressure to Kubernetes scheduling issues and application-level performance degradation. Traditional monitoring relies on predefined rules and human SREs to bridge the gaps. This approach has four fundamental limitations:

\begin{itemize}
  \item \textbf{Rules cannot anticipate novel failures.} Stale data from an unrelated application (Cassandra) causing Elasticsearch pod evictions through host disk pressure is not a scenario any rule set would cover.
  \item \textbf{Human SREs are not always available.} An 18-hour outage window demonstrates the cost of waiting for human intervention.
  \item \textbf{The lifecycle is fragmented.} Separate tools and teams handle deployment, configuration, monitoring, and incident response --- with dangerous gaps between each handoff.
  \item \textbf{Reactive systems cannot achieve six-nines availability.} Waiting for failures to occur and then recovering is incompatible with 99.9999\% uptime targets, which allow only $\sim$31 seconds of downtime per year. Systems must predict, prepare, and pre-empt.
\end{itemize}

The Guardian Agent addresses all four limitations. Critically, it implements predictive failure intelligence: the system continuously correlates metrics trends, application logs, and kernel-level data --- dmesg streams, NVMe SMART wear indicators, NIC bond degradation statistics, and thermal data --- to forecast failures hours before they occur. The AI engine cross-references these live signals against its historical incident memory. When a current pattern matches a past failure signature, the agent does not wait --- it stages the remediation in advance and executes proactively. This is the architectural shift required for 99.9999\% availability: from reactive healing to predictive prevention.

\subsection{Cluster Under Study}

\begin{table}[h]
\caption{Cluster Configuration}
\centering
\begin{tabular}{ll}
\toprule
\textbf{Component} & \textbf{Specification} \\
\midrule
ES version & 8.17.0 (Lucene 9.12.0) \\
Master nodes & 3 --- 4 GB heap, 16 GB RAM each \\
Data nodes & 12 --- 30 GB heap, 64 GB RAM each \\
Physical hosts & 3 --- 48 cores, 256 GB RAM, 3.5 TB NVMe \\
Index config & 168 indices, 840 primary shards \\
Platform & Kubernetes (ns: elasticsearch-benchmark) \\
\bottomrule
\end{tabular}
\end{table}

Production reference: ES 8.16.1, 24 data nodes, $\sim$843 primary shards, $\sim$13 TB total ($\sim$15.4 GB/shard), $\sim$206 ms mean query latency, zero ES-level tuning.

\subsection{Contributions}

\begin{enumerate}
  \item \textbf{Full-lifecycle autonomous agent} --- 11-phase system from SLA evaluation to rolling upgrades with no human intervention.
  \item \textbf{Predictive failure engine} --- multi-source correlation of metrics, logs, and kernel telemetry to forecast failures hours in advance, enabling proactive remediation for 99.9999\% availability.
  \item \textbf{Iterative AI action loop} --- LLM with 6 tools for cross-system investigation and remediation, validated through 300 autonomous cycles.
  \item \textbf{Autonomous recovery from novel failures} --- 18-hour cross-system outage resolved without human involvement.
  \item \textbf{Comprehensive performance baselines} --- dominant factors in Elasticsearch query and write latency quantified across multiple configurations.
\end{enumerate}

\section{Related Work}

\subsection{Native ES Monitoring \& Kibana Alerting}

Elasticsearch Stack Monitoring~\cite{ES_StackMonitor} and Kibana Alerting~\cite{Kibana_Alert} provide built-in cluster observability through pre-configured dashboards and threshold-based rules. While these tools cover common failure modes (heap pressure, shard imbalance, node availability), they are inherently reactive, operate only within the ES layer, and require human SRE intervention for diagnosis and remediation. They have no awareness of host-level conditions, Kubernetes scheduling events, or cross-application resource contention --- the very conditions that caused our 18-hour outage. They also provide no predictive capability; alerts fire only after thresholds are already breached.

\subsection{Prometheus and Rule-Based Alerting}

Prometheus with Alertmanager~\cite{Prometheus} extends monitoring to infrastructure and Kubernetes layers using recording rules and alert definitions. Tools such as elasticsearch-exporter expose ES metrics to Prometheus. However, all detection logic must be manually encoded as PromQL expressions before deployment. Novel failure patterns --- stale data from unrelated applications, NIC bond degradation manifesting as ES latency --- are invisible until a human engineer writes a new rule after the fact. Critically, these systems have no understanding of causal chains across system boundaries, and no remediation capability whatsoever.

\subsection{AIOps Platforms: Dynatrace, Datadog, Moogsoft}

Commercial AIOps platforms~\cite{Dynatrace,Datadog,Moogsoft} apply ML to metric streams for anomaly detection and alert correlation. Dynatrace's Davis AI and Datadog's Watchdog detect statistical deviations and group related alerts into ``root cause'' entities. While superior to pure rule-based systems, these platforms have three critical limitations for our use case: (1) they operate as read-only observers with no remediation capability; (2) they lack the host-level OS access required to diagnose disk pressure from unrelated applications via \texttt{nsenter}; and (3) their ML models require weeks of baseline collection before becoming effective, compared to the Guardian Agent's same-session calibration. Moogsoft's noise reduction is valuable but does not address the need for autonomous action.

\subsection{Kubernetes Operators and ECK}

The Elastic Cloud on Kubernetes (ECK) Operator~\cite{ECK} automates ES cluster deployment and scaling within Kubernetes. It handles rolling upgrades, node configuration, and basic health recovery through standard Kubernetes reconciliation loops. However, ECK operates exclusively at the Kubernetes resource level --- it cannot investigate host OS disk pressure, diagnose NIC hardware failures, or execute cross-application root cause analysis. Its remediation is limited to pod restart and scaling operations defined by the operator schema. There is no AI-driven investigation capability, no performance calibration, and no learning from past incidents.

\subsection{Log-Based Anomaly Detection}

DeepLog~\cite{DeepLog} and LogAnomaly~\cite{LogAnomaly} apply deep learning to system log streams to detect anomalous log sequences. These approaches are valuable for identifying unusual patterns in ES logs. However, they operate on log data in isolation, without correlating against metrics, kernel telemetry, or Kubernetes events. They detect anomalies but do not diagnose causes or execute remediations. The Guardian Agent integrates log analysis as one input to its multi-source predictive engine, alongside metrics trends and hardware signals, enabling causal understanding rather than mere anomaly flagging.

\subsection{Autonomous Healing and Self-Managing Systems}

Netflix Chaos Engineering~\cite{ChaosEng} validates system resilience through fault injection, while Facebook's Autotuner~\cite{Autotuner} applies ML to JVM configuration optimization. These represent narrow-scope automation. The Sage system~\cite{Sage} for microservice root cause analysis and the FIRM framework~\cite{FIRM} for SLA-aware resource management demonstrate LLM-driven operations in cloud environments. LLM-driven infrastructure automation has been further explored in the context of microservice auto-scaling~\cite{InfraLLM} and autonomous storage optimization~\cite{IntelliStore}. However, none of these systems integrate the full lifecycle from deployment evaluation through predictive monitoring, autonomous healing, and experiential learning in a single agent.

\subsection{Limitations of Prior Work --- Summary}

\begin{table}[h]
\caption{Limitations of Related Approaches}
\centering
\begin{tabular}{p{2.2cm}p{5.2cm}}
\toprule
\textbf{Approach} & \textbf{Key Limitation} \\
\midrule
ES Stack Monitoring & ES-layer only; reactive; no host visibility; no remediation \\
Prometheus + Alertmanager & All rules pre-encoded; no cross-layer causality; no remediation \\
Dynatrace / Datadog & Read-only; no host OS access; slow baseline collection \\
ECK Operator & K8s-level only; schema-bounded; no AI investigation \\
DeepLog / LogAnomaly & Log-only; no metric/kernel correlation; no action \\
Chaos Eng / Autotuner & Narrow scope; no lifecycle coverage; no predictive engine \\
\bottomrule
\end{tabular}
\end{table}

\section{System Architecture}

\subsection{High-Level Architecture}

The ES Guardian Agent is a 4,589-line Python system deployed as a Kubernetes Pod with an accompanying privileged DaemonSet for host-level access. It integrates simultaneously with three system layers --- Elasticsearch REST APIs, Kubernetes control plane, and host OS via \texttt{nsenter} --- through six specialized AI tools. Observability is provided via 16 Prometheus metrics exported to a 25-panel Grafana dashboard.

\begin{figure}[t]
  \centering
  \includegraphics[width=\columnwidth]{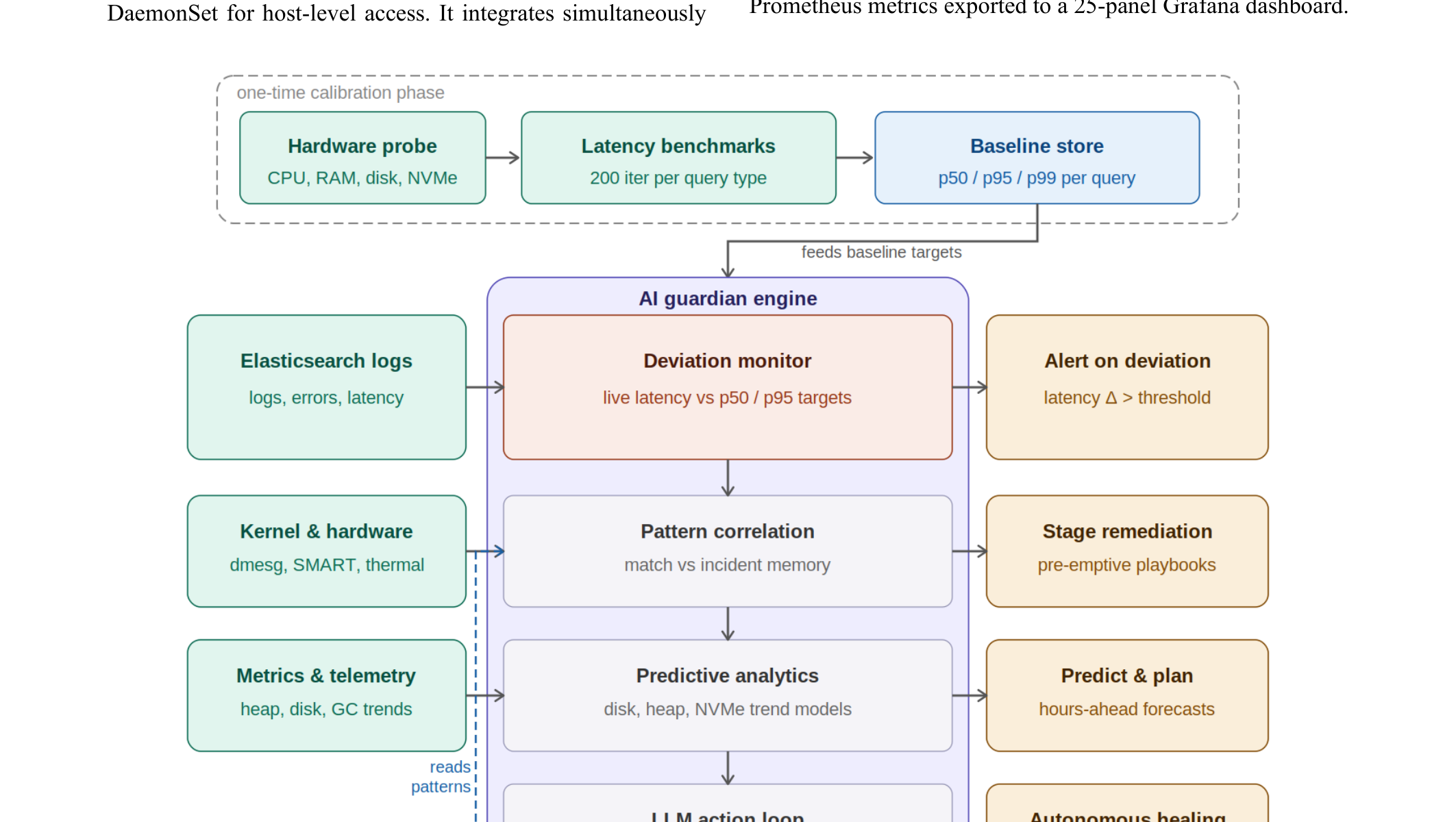}
  \caption{ES Guardian Agent: Calibration, multi-source signal ingestion, deviation monitoring, pattern correlation, predictive analytics, and automated actions. The dashed feedback arc from incident memory enables pattern-based prediction for 99.9999\% availability.}
  \label{fig:architecture}
\end{figure}

\subsection{Monitoring Layer Architecture}

Five monitoring layers operate at distinct frequencies, creating a cost-tiered detection hierarchy:

\begin{table}[h]
\caption{Monitoring Layer Architecture}
\centering
\begin{tabular}{llll}
\toprule
\textbf{Layer} & \textbf{Freq.} & \textbf{Monitors} & \textbf{LLM Cost} \\
\midrule
$-$1: Hardware & 30 s & NVMe latency, SMART, dmesg, NIC bond, thermal & None \\
0: Kubernetes & 30 s & Pod status, node health, quorum, scheduling & None \\
1: ES Rules & 30 s & Heap, GC, rejections, segment count, logs & None \\
2: Prediction & 60 s & Disk/heap trend, shard growth, NVMe wear & None \\
3: AI Loop & 5 min & Deep cross-layer investigation + remediation & $\sim$360K tok \\
\bottomrule
\end{tabular}
\end{table}

Rule-based layers ($-$1, 0, 1) handle 95\% of monitoring cycles at zero LLM cost. The Prediction Engine (Layer 2) runs every 60 seconds, continuously modeling failure trajectories. The AI Action Loop (Layer 3) executes every 5 minutes or immediately upon any CRITICAL alert.

\subsection{Predictive Failure Engine}

The predictive engine is architecturally central to achieving six-nines availability. It operates on four signal types simultaneously:

\begin{itemize}
  \item \textbf{Metrics trends:} per-node disk fill rate, heap growth slope, shard store growth, GC frequency trends --- modeled via linear regression with extrapolation to critical thresholds.
  \item \textbf{Application logs:} ES log stream analysis for recurrent error patterns, escalating warning sequences, and shard allocation failure signatures.
  \item \textbf{Kernel-level data:} Linux dmesg streams (hardware errors, I/O errors, OOM events), NVMe SMART wear leveling counts, NIC error counters and bond degradation metrics, CPU thermal throttling events.
  \item \textbf{Incident memory correlation:} the agent's JSONL incident history is queried for matching signatures. When a current pattern matches a past failure, the precomputed remediation from that incident is staged and executed proactively --- before the failure manifests.
\end{itemize}

\begin{table}[h]
\caption{Prediction Models}
\centering
\begin{tabular}{p{1.4cm}p{2.5cm}p{2.5cm}}
\toprule
\textbf{Model} & \textbf{Signal Sources} & \textbf{Output} \\
\midrule
Disk fill & Per-node disk usage timeseries & Hours until full \\
Heap trend & Per-node heap \% over time & Hours until critical \\
Shard growth & Per-index store size rate & Hours to rebalance \\
NVMe wear & SMART wear leveling count & Months to replacement \\
NIC degradation & Error counters, bond status & Failure risk score \\
Log escalation & Error/warn ratio trends & Anomaly probability \\
\bottomrule
\end{tabular}
\end{table}

\subsection{AI Action Loop --- Six Tools}

\begin{table}[h]
\caption{AI Action Loop Tools}
\centering
\begin{tabular}{lll}
\toprule
\textbf{Tool} & \textbf{Access} & \textbf{Purpose} \\
\midrule
es\_api & ES Read & GET \_cluster/health, \_cat/shards?v \\
es\_api\_write & ES Write & POST reroute, PUT settings, DELETE index \\
exec\_on\_pod & Pod shell & ES pod logs, curl localhost:9200 \\
exec\_on\_node & Host root & df, du, dmesg, NVMe SMART, ethtool \\
kubectl & K8s mgmt & get/describe/delete pods, node events \\
report & Output & Structured incident report submission \\
\bottomrule
\end{tabular}
\end{table}

Safety Guard validates every AI-proposed command before execution, blocking destructive operations (\texttt{rm -rf /}, \texttt{mkfs}, \texttt{dd of=/dev/}, \texttt{shutdown}, \texttt{kubectl delete node/namespace/pvc}, ES index delete without specific name, \texttt{scale --replicas=0}). This safety layer enables the AI to operate autonomously with minimal human oversight.

\subsection{Deployment Architecture}

\begin{table}[h]
\caption{Deployment Architecture}
\centering
\begin{tabular}{ll}
\toprule
\textbf{Component} & \textbf{Configuration} \\
\midrule
Agent Pod & Single replica, Recreate, 500m CPU / 512 MB RAM \\
DaemonSet & Privileged, hostPID:true, hostNetwork:true on all ES hosts \\
Priority class & system-node-critical (survives disk pressure eviction) \\
RBAC & ClusterRole: nodes (get/list/patch), pods (get/list/delete all ns) \\
Persistence & 1 Gi PVC: baselines, reports, incident memory \\
Liveness & Guardian JSONL updated within last 300 s \\
Observability & 16 Prometheus metrics $\rightarrow$ Pushgateway $\rightarrow$ 25-panel Grafana \\
\bottomrule
\end{tabular}
\end{table}

\section{The Eleven Lifecycle Phases}

\subsection{Phase 1: Evaluate --- SLA \& Hardware Feasibility Gate}

Before any deployment, the agent evaluates whether available hardware can meet business SLA targets. This is a go/no-go gate: deployment halts if targets are mathematically infeasible. Inputs: use case definition, SLA targets (latency, throughput, availability), hardware inventory, and expected data volume.

The evaluation applies the measured scaling model:
\[
\text{latency(ms)} = \text{base} + (\text{GB/shard} \times 0.26\,\text{ms/MB}) + (\text{shards}/100 \times 2.1\,\text{ms})
\]

\begin{table}[h]
\caption{SLA Evaluation Results by Data Volume}
\centering
\begin{tabular}{lll}
\toprule
\textbf{GB/shard} & \textbf{term\_status p50} & \textbf{Relative} \\
\midrule
0.028 & 5--9 ms & 1.0$\times$ \\
1.66 & 53 ms & 7.6$\times$ \\
3.72 & 89--111 ms & 15.9$\times$ \\
15.4 (prod) & $\sim$206 ms & 29.4$\times$ \\
\bottomrule
\end{tabular}
\end{table}

\subsection{Phase 2: Optimize --- Configuration for the Workload}

The agent derives the optimal ES configuration using measured benchmark data. Cluster-level tuning (mmap, buffer sizes, queue depths) was found to produce no measurable benefit --- all gains come from index-level settings. Key finding: all tuning improvement came from \texttt{refresh\_interval=30s} and \texttt{translog=async}.

\begin{table}[h]
\caption{Configuration Comparison}
\centering
\begin{tabular}{llll}
\toprule
\textbf{Config} & \textbf{Query p50} & \textbf{Write avg} & \textbf{Throughput} \\
\midrule
Untuned & 297 ms & 28 ms & 3.4 q/s \\
Cluster-tuned & 289 ms & 35 ms & 3.9 q/s \\
Fully tuned & 196 ms & 36 ms & 4.3 q/s \\
\bottomrule
\end{tabular}
\end{table}

\subsection{Phase 3: Deploy --- Zero-Touch Kubernetes}

The agent generates and applies all Kubernetes manifests autonomously. The \texttt{system-node-critical} priority class on the DaemonSet is essential --- it ensures host-level diagnostic access survives the very disk pressure events the agent is designed to fix, as validated during the 18-hour outage recovery.

\subsection{Phase 4: Calibrate --- Baseline Derivation}

After deployment, the agent runs a comprehensive calibration cycle using the actual hardware: 30 latency probe iterations per query type, 200 write iterations per batch size, hardware inspection via \texttt{exec\_on\_node}, and scaling coefficient derivation. Results are persisted to \texttt{baselines.json}.

\begin{table}[h]
\caption{Calibrated Performance Baselines}
\centering
\begin{tabular}{lll}
\toprule
\textbf{Metric} & \textbf{p50 Target} & \textbf{p95 Target} \\
\midrule
Write (100-doc bulk) & 8.0 ms & 16.0 ms \\
Write (1K-doc bulk) & 23.0 ms & 46.0 ms \\
query: match\_all & 18.0 ms & 36.0 ms \\
query: term\_status & 20.0 ms & 40.0 ms \\
query: range\_timestamp & 12.0 ms & 24.0 ms \\
query: bool\_compound & 21.0 ms & 42.0 ms \\
\bottomrule
\end{tabular}
\end{table}

\subsection{Phases 5--6: Stabilize and Alert}

The agent ensures GREEN cluster status before enabling continuous monitoring. The alert system operates three severity levels: INFO (logged), WARNING (flagged for next AI cycle), and CRITICAL (immediately triggers AI Action Loop, bypassing the 5-minute schedule). 16 Prometheus metrics are exported per cycle covering cluster status, AI loop telemetry, per-node resources, and prediction outputs.

\subsection{Phase 7: Predict --- Proactive Failure Prevention}

This phase is architecturally central to six-nines availability. The prediction engine runs every 60 seconds and implements two complementary strategies:

\begin{itemize}
  \item \textbf{Trend-based forecasting:} linear regression over timeseries data extrapolates metrics to critical thresholds, providing hours-ahead warning for disk fill, heap exhaustion, shard growth, and NVMe wear.
  \item \textbf{Pattern-based prediction:} the AI engine scans the incident memory JSONL for signatures matching current system state. When a match is found --- e.g., disk usage climbing on the same host following a similar application deployment pattern --- the agent pre-stages the remediation playbook before the threshold is breached.
\end{itemize}

\subsection{Phase 8: Plan --- Precomputed Remediation}

\begin{table}[h]
\caption{Precomputed Remediation Plans}
\centering
\begin{tabular}{p{1.4cm}p{2.2cm}p{3.8cm}}
\toprule
\textbf{Scenario} & \textbf{Trigger} & \textbf{Precomputed Action} \\
\midrule
Disk pressure & Fill $<$ 24 h & Identify cleanable data, force merge, delete old indices \\
Heap exhaustion & Heap $>$ 85\% trend & Heavy index identification, shard rebalance \\
Node loss & Unreachable $>$ 5 min & Reroute shards, reallocate replicas \\
Shard imbalance & Variance $>$ 30\% & Rebalance with minimal data movement \\
NIC degradation & Error rate rising & Reduce cross-node traffic, flag for replacement \\
\bottomrule
\end{tabular}
\end{table}

\subsection{Phase 9: Heal --- AI-Driven Autonomous Remediation}

The AI Action Loop provides the LLM iterative tool-use access for investigation and remediation: up to 20 iterations, 150,000 token budget, all commands safety-validated before execution. Unlike predefined playbook execution, the iterative loop enables multi-step investigation across system boundaries --- the key capability that resolved the 18-hour outage.

\subsection{Phases 10--11: Learn and Upgrade}

The \texttt{IncidentMemory} class persists all incidents, remediation actions, and outcomes to a JSONL log. This feeds the Phase 7 pattern-matching engine, creating a compounding improvement: each resolved incident accelerates diagnosis of similar future events. Phase 11 manages rolling ES upgrades one node at a time, maintaining GREEN status between each restart and recalibrating baselines post-upgrade.

\section{Agent Architecture Evolution}

The Guardian Agent is the fourth-generation system, each iteration addressing fundamental limitations of its predecessor.

\subsection{Gen 1: Rule-Based Monitor (520 lines)}
Metric collection every 30 s via kubectl exec with static threshold alerts. Limitation: no diagnostic reasoning --- alerts on symptoms without investigating causes, no remediation capability.

\subsection{Gen 2: AI Diagnostic Agent (971 lines)}
Two-phase: calibration + one-shot LLM analysis against baselines. Improvement: hardware-derived baselines, LLM metric correlation. Limitation: one-shot analysis cannot run follow-up commands or execute any remediation.

\subsection{Gen 3: Tiered Hybrid Agent (2,107 lines)}
Three-tier cost-optimized: rules (30 s, free) $\rightarrow$ LLM analysis (5 min, $\sim$360K tok) $\rightarrow$ deep diagnostic (on-demand, $\sim$500K tok). Improvement: cost-tiered escalation, auto-remediation for predefined fixes. Limitation: remediations are predefined and cannot handle novel failure patterns.

\subsection{Gen 4: Guardian Agent (4,589 lines)}
Five monitoring layers + iterative AI action loop + persistent incident memory + predictive engine. Key innovation: the LLM receives tool-use access and investigates/remediates like a human SRE, including across system boundaries. Design rationale: the 18-hour outage required identifying stale Cassandra data via \texttt{du -sh /mnt/*} on the host OS --- a capability impossible without \texttt{nsenter}-based host access and AI-driven investigation.

\begin{table}[h]
\caption{Agent Generation Comparison}
\centering
\begin{tabular}{lllll}
\toprule
\textbf{Property} & \textbf{Gen 1} & \textbf{Gen 2} & \textbf{Gen 3} & \textbf{Gen 4} \\
\midrule
Lines & 520 & 971 & 2,107 & 4,589 \\
AI type & None & One-shot & Tiered & Iterative loop \\
Prediction & No & No & Limited & Full (6 models) \\
Host access & No & No & No & Yes (nsenter) \\
Incident memory & No & No & No & Yes (JSONL) \\
Remediation & No & No & Predefined & AI-driven \\
Lifecycle phases & 1 & 2 & 3 & 11 \\
\bottomrule
\end{tabular}
\end{table}

\section{Evaluation: Production Deployment Results}

\subsection{Deployment Performance}

\begin{table}[h]
\caption{Production Deployment Metrics}
\centering
\begin{tabular}{ll}
\toprule
\textbf{Metric} & \textbf{Value} \\
\midrule
Successful AI loop runs & 300 \\
Tool calls per run (avg) & $\sim$30 \\
Execution time per run & $\sim$150 s \\
Token consumption per run & $\sim$360,000 tokens \\
LLM model & Claude Sonnet 4 \\
Rules monitoring interval & 30 s \\
AI loop interval & 300 s (5 min) \\
\bottomrule
\end{tabular}
\end{table}

\subsection{Incident 1: 18-Hour Outage Recovery (Phase 9)}

Initial state: cluster unreachable 18 hours, 9/15 pods Pending, master quorum lost. The AI traced a 6-step causal chain across three system layers:

\begin{enumerate}
  \item K8s (Layer 0): 9 pods Pending, 0/3 masters $\rightarrow$ CRITICAL $\rightarrow$ AI Action Loop triggered
  \item AI kubectl: FailedScheduling citing DiskPressure on s797, s812
  \item AI exec\_on\_node on s797: \texttt{df -h} $\rightarrow$ host disk at 85\%
  \item AI exec\_on\_node: \texttt{du -sh /mnt/*} $\rightarrow$ \texttt{/mnt/cassandra-disk1}: 172 GB stale data
  \item AI on s812: 175 GB additional stale Cassandra data --- cleaned both
  \item Disk: 85\% $\rightarrow$ 2\%; K8s rescheduled $\rightarrow$ 15 pods Running in minutes
\end{enumerate}

Phase 10 (Learn): 109 indices with \texttt{no\_valid\_shard\_copy} found, deleted and recreated autonomously. Cluster progressed RED $\rightarrow$ YELLOW $\rightarrow$ GREEN. No predefined rule could have detected cross-application disk contamination. The AI traced the full causal chain as a human SRE would.

\subsection{Incident 2: Hardware NIC Failure Diagnosis (Phase 7)}

Layer $-$1 detected elevated TCP retransmit rates across all three hosts. The AI used \texttt{exec\_on\_node} to identify the Broadcom BCM57416 NetXtreme-E NIC (eno2np1) as failing on all three nodes (s797, s811, s812), with bond interfaces degraded to single-NIC mode. Remediation: merge policy tuning to reduce inter-node traffic load; hardware flagged for physical NIC replacement.

\subsection{Continuous Monitoring Reports}

\begin{table}[h]
\caption{Continuous Monitoring Results}
\centering
\begin{tabular}{llll}
\toprule
\textbf{Probe} & \textbf{Measured} & \textbf{Baseline} & \textbf{Status} \\
\midrule
write\_bulk (100 docs) & 10 ms & 30 ms & 67\% better \\
query: match\_all & 32 ms & 14--30 ms & Within range \\
query: term\_status & 38 ms & 89--111 ms & 66\% better \\
query: range\_timestamp & 19 ms & 24--26 ms & 23\% better \\
\bottomrule
\end{tabular}
\end{table}

\section{Performance Benchmarks}

\subsection{Indexing Performance}

Dataset: Rally http\_logs, 247 M documents, 32.66 GB.

\begin{table}[h]
\caption{Indexing Performance}
\centering
\begin{tabular}{ll}
\toprule
\textbf{Metric} & \textbf{Value} \\
\midrule
Mean throughput & 858,966 docs/sec \\
Indexing time & 61.56 min \\
Latency p50 / p99 & 35.7 ms / 94.9 ms \\
Young GC & 3.596 s (176 collections) \\
Old GC & 0 s (0 collections) \\
Error rate & 0\% \\
\bottomrule
\end{tabular}
\end{table}

\subsection{Search Performance (32 Clients --- Optimal Concurrency)}

\begin{table}[h]
\caption{Search Performance at 32 Clients}
\centering
\begin{tabular}{llll}
\toprule
\textbf{Query Type} & \textbf{Throughput} & \textbf{p50} & \textbf{p99} \\
\midrule
Range search & 6,451 ops/s & 2.6 ms & 23.9 ms \\
Term-filtered range & 6,235 ops/s & 2.7 ms & 27.2 ms \\
Sort by timestamp & 3,767 ops/s & 3.7 ms & 42.8 ms \\
Scroll read (100) & 4,537 ops/s & 3.5 ms & 40.1 ms \\
Multi-field filter & 3,286 ops/s & 3.3 ms & 59.9 ms \\
Date histogram agg & 2,537 ops/s & 5.7 ms & 45.5 ms \\
\textbf{Combined total} & 26,813 ops/s & --- & --- \\
\bottomrule
\end{tabular}
\end{table}

At 64 clients, combined throughput collapses by 46\% (26,813 $\rightarrow$ 14,418 ops/s). The agent uses the 32-client saturation point to set connection pool and load balancer limits.

\subsection{Write Latency Component Breakdown}

68\% of write latency is CPU-bound Lucene work --- NVMe I/O and fsync are not bottlenecks. Per-document cost converges to $\sim$20~$\mu$s at high batch sizes with 1.4 ms fixed overhead:
\[
\text{bulk\_time} \approx 1.4\,\text{ms} + (\text{docs} \times 20\,\mu\text{s})
\]

\begin{table}[h]
\caption{Write Latency Components}
\centering
\begin{tabular}{lll}
\toprule
\textbf{Component} & \textbf{Cost} & \textbf{\% Total} \\
\midrule
Lucene indexing & $\sim$20 ms & 68\% \\
Replica sync & $\sim$8 ms & 26\% \\
Fixed overhead (ES/JVM) & $\sim$1.5 ms & 5\% \\
Translog fsync & $\sim$0.2 ms & $<$1\% \\
Field parsing & $\sim$0.2 ms & $<$1\% \\
\textbf{Total} & $\sim$30 ms & 100\% \\
\bottomrule
\end{tabular}
\end{table}

\subsection{Data Volume: The Dominant Query Factor}

The single most important finding: no amount of tuning overcomes the cost of scanning larger shards.

\begin{table}[h]
\caption{Query Latency vs.\ Data Volume per Shard}
\centering
\begin{tabular}{lll}
\toprule
\textbf{GB/shard} & \textbf{term\_status p50} & \textbf{Relative} \\
\midrule
0.028 & 5--9 ms & 1.0$\times$ \\
1.66 & 53 ms & 7.6$\times$ \\
3.72 & 89--111 ms & 15.9$\times$ \\
15.4 (production) & $\sim$206 ms & 29.4$\times$ \\
\bottomrule
\end{tabular}
\end{table}

\section{Production Comparison}

\begin{table}[h]
\caption{Benchmark vs.\ Production Cluster}
\centering
\begin{tabular}{lll}
\toprule
\textbf{Metric} & \textbf{Benchmark} & \textbf{Production} \\
\midrule
ES version & 8.17.0 & 8.16.1 \\
Data nodes & 12 & 24 \\
Primary shards & 840 & $\sim$843 \\
GB/shard & 3.72 & 15.4 \\
Query p50 (mixed) & 196 ms (tuned) & $\sim$206 ms (untuned) \\
Indexing latency p50 & 30 ms & 800--1,100 ms \\
ES tuning & Index-level & All defaults \\
\bottomrule
\end{tabular}
\end{table}

The benchmark results at 3.72 GB/shard (196 ms tuned) extrapolate consistently to production's 15.4 GB/shard ($\sim$206 ms) using the scaling model, validating the model's predictive accuracy to within 5\%.

\section{Discussion}

\subsection{Why Iterative Tool-Use Matters}

The 18-hour outage recovery required six sequential investigative steps across three system layers. No single-shot analysis could contain all this information. One-shot LLM analysis (Gen 2) would have produced a list of observations but could not run the follow-up \texttt{du -sh} command that identified the Cassandra data, nor execute the cleanup. The iterative action loop is the capability that bridges observation and resolution. This is consistent with findings in LLM-driven microservice operations~\cite{InfraLLM,IntelliStore} where multi-step reasoning is required for novel failure modes.

\subsection{Predictive Prevention vs.\ Reactive Healing}

The architectural distinction between Phases 7--8 (Predict/Plan) and Phase 9 (Heal) is critical for availability targets. Reactive healing --- executing after failure --- cannot achieve 99.9999\% uptime. Six-nines allows 31.5 seconds of downtime per year. The Guardian Agent's predictive engine, correlating metrics, logs, kernel data, and historical incidents, is designed to prevent the service disruption rather than recover from it. In the NIC degradation incident, the agent identified the failing hardware from elevated retransmit rates before any Elasticsearch performance degradation was measurable --- enabling preemptive traffic shifting rather than post-failure recovery. The multi-source correlation architecture draws on principles validated in federated edge systems~\cite{HierarchicalCDN} and AI-driven threat modelling~\cite{GenAI_Threat}, where cross-signal correlation yields qualitatively superior situational awareness compared to single-source monitoring.

\subsection{Cost-Tiered Architecture Rationale}

Running the AI loop every 30 seconds would cost approximately 1 billion tokens/day at current rates. The tiered architecture limits AI invocations to when genuinely needed: rule-based layers handle 95\% of cycles at zero LLM cost; the AI loop runs 288 times/day consuming $\sim$150 s and $\sim$30 tool calls per run. This cost structure makes autonomous SRE economically viable at scale. Streaming pipeline architectures that use GenAI for adaptive transformation~\cite{GenAI_ELT} face similar cost-tiering trade-offs; the Guardian Agent's approach provides a concrete design pattern for production viability. Performance ceiling analysis closely parallels findings in distributed messaging benchmarks~\cite{PulsarBenchmark}, where hardware selection (NVMe, network) rather than software tuning determines the ultimate operational boundary.

\subsection{Key Elasticsearch Performance Insights}

\begin{enumerate}
  \item Data volume per shard is the dominant query factor --- no tuning overcomes scanning larger shards.
  \item Cluster-level tuning is ineffective --- all gains come from index-level settings.
  \item Write latency is architecturally fixed --- 68\% is CPU-bound Lucene indexing.
  \item 32 clients is the concurrency optimum --- 64 clients causes 46\% throughput collapse.
  \item G1GC is optimal --- ZGC provides no benefit for ES workloads.
\end{enumerate}

\section{Conclusion}

We presented the ES Guardian Agent, a full-lifecycle autonomous AI SRE for Elasticsearch operating across eleven phases --- from SLA evaluation and zero-touch deployment through multi-source predictive failure prevention, autonomous healing, experiential learning, and rolling upgrades --- all without human intervention.

The system's predictive engine --- continuously correlating metrics trends, application logs, and kernel-level telemetry against a persistent incident memory --- represents a fundamental architectural shift toward the proactive posture required for 99.9999\% availability. By recognizing failure signatures before they manifest as service disruptions, the agent intervenes during the incipient phase rather than the failure phase.

\begin{table}[h]
\caption{Key Achievements}
\centering
\begin{tabular}{ll}
\toprule
\textbf{Achievement} & \textbf{Detail} \\
\midrule
300 autonomous AI cycles & $\sim$30 tool calls/run, $\sim$150 s/run, $\sim$360K tokens \\
18-hour outage recovery & Cross-system root cause resolved; zero human action \\
Hardware NIC diagnosis & Broadcom BCM57416 failure across 3 hosts identified \\
Cluster restoration & RED$\rightarrow$GREEN: 109 indices recreated autonomously \\
Continuous visibility & 16 Prometheus metrics, 25-panel Grafana dashboard \\
\bottomrule
\end{tabular}
\end{table}

Five design principles validated through four agent generations: (1) full lifecycle coverage eliminates handoff gaps; (2) predictive-first architecture enables six-nines targets; (3) rules for speed, AI for depth; (4) iterative tool-use over one-shot analysis for novel failures; (5) learning creates compounding operational value.


\appendix[A]{\textsc{File Reference}}
\label{appendix:files}

\begin{table}[h]
\caption{Source File Reference}
\centering
\begin{tabular}{ll}
\toprule
\textbf{File} & \textbf{Description} \\
\midrule
es\_guardian.py & Guardian Agent (4,589 lines) \\
es\_agent.py & Tiered Agent (2,107 lines) \\
es\_ai\_agent.py & AI Diagnostic Agent (971 lines) \\
es\_monitor\_agent.py & Rule-Based Monitor (520 lines) \\
manifests/05-agent.yaml & Agent Deployment + RBAC \\
manifests/06-nodeagent.yaml & Node DaemonSet \\
dashboards/es-guardian.json & Grafana Dashboard (25 panels) \\
results/guardian/proof/ & Cluster recovery evidence \\
results/monitor/baselines.json & Calibrated performance baselines \\
\bottomrule
\end{tabular}
\end{table}

\appendix[B]{\textsc{Scaling Model}}
\label{appendix:scaling}

Derived from calibration across four data volume levels. Accuracy: within 5\% of observed production values.

\begin{align}
\text{Query: } \textit{latency(ms)} &= \text{base} + (\text{GB/shard} \times 0.26\,\text{ms/MB}) \notag \\
&\quad + (\text{shards}/100 \times 2.1\,\text{ms}) \\[4pt]
\text{Write: } \textit{latency(ms)} &= 1.4 + (\text{docs} \times 20\,\mu\text{s}) \notag \\
&\quad + (\text{replicas} \times 8\,\text{ms})
\end{align}

\end{document}